\documentclass[journal]{IEEEtran}
\usepackage{amsmath,amsfonts,amssymb}
\usepackage{algorithmic}
\usepackage{array}
\usepackage[caption=false,font=normalsize,labelfont=sf,textfont=sf]{subfig}
\usepackage{textcomp}
\usepackage{stfloats}
\usepackage{url}
\usepackage{verbatim}
\usepackage{graphicx}
\usepackage[table]{xcolor}
\usepackage{xcolor}
\usepackage{xfp}
\usepackage{amsmath}
\usepackage{tikz}
\usepackage{pgfplots}
\usepackage{contour}
\usepackage{mathtools}
\usepackage[english]{babel}
\usepackage{orcidlink}
\usepackage{pgfplots,pgfplotstable}
\usepackage[noadjust]{cite}
\usepackage{float}
\usepackage{environ}

\pgfplotsset{compat=1.18}
\usepgfplotslibrary{colormaps}
\usepgfplotslibrary{colorbrewer}
\usepgfplotslibrary{groupplots}
\usetikzlibrary{fpu}
\usetikzlibrary{shapes.geometric}

\def\BibTeX{{\rm B\kern-.05em{\sc i\kern-.025em b}\kern-.08em
    T\kern-.1667em\lower.7ex\hbox{E}\kern-.125emX}}
\usepackage{balance}

\newcommand{\Aeff}{\ensuremath{A_\mathrm{eff}}}

\DeclareMathOperator*{\sinc}{\mathrm{sinc}}

\colorlet{secondarytext}{gray}

\definecolor{grad-1}{HTML}{C7B3CC}
\definecolor{grad-2}{HTML}{A7ABC7}
\definecolor{grad-3}{HTML}{87A3C2}
\definecolor{grad-4}{HTML}{669ABC}
\definecolor{grad-5}{HTML}{4692B7}
\definecolor{grad-6}{HTML}{268AB2}

\NewEnviron{splitfit}[1][\linewidth]{\resizebox{#1}{!}{$\begin{split}\BODY\end{split}$}}

\begin{document}
\title{Optimising O-to-U Band Transmission Using Fast ISRS Gaussian Noise Numerical Integral Model}
\author{
    Mindaugas~Jarmolovi\v{c}ius~\orcidlink{0000-0002-0456-110X}, Daniel~Semrau, Henrique~Buglia~\orcidlink{0000-0003-1634-0926}, Mykyta~Shevchenko~\orcidlink{0000-0001-7094-1322}, Filipe~M.~Ferreira~\orcidlink{0000-0002-1533-843X}, Eric~Sillekens~\orcidlink{0000-0003-1032-6760}, Polina~Bayvel~\orcidlink{0000-0003-4880-3366} and Robert~I.~Killey\thanks{M.~Jarmolovi\v{c}ius, H.~Buglia, Filipe~M.~Ferreira, E.~Sillekens, P.~Bayvel and R.~I.~Killey are with Optical Networks Group, Department of Electronic and Electrical Engineering, UCL (University College London), London, UK}\thanks{D.~Semrau is with Infinera Corporation, 9005 Junction Drive, Annapolis Junction, MD 20701, USA}\thanks{M.~Shevchenko is with National Physical Laboratory (NPL), Hampton Road, Teddington, Middlesex, TW11 0LW, UK}
}

\markboth{Journal of XXXX,~Vol.~XX, No.~XX, XX~XXXX}{XXXX \MakeLowercase{\textit{et al.}}}

\maketitle

\begin{abstract}
We model the transmission of ultrawideband signals, including wavelength-dependent fibre parameters: dispersion, nonlinear coefficient and effective fibre core area. To that end, the inter-channel stimulated Raman scattering Gaussian noise integral model is extended to include these parameters. The integrals involved in this frequency-domain model are numerically solved in hyperbolic coordinates using a Riemann sum. The model implementation is designed to work on parallel GPUs and is optimised for fast computational time. The model is valid for Gaussian-distributed signals and is compared with the split-step Fourier method, for transmission over standard single-mode fibre (SSMF) in the O-band (wavelengths around the zero-dispersion wavelength), showing reasonable agreement. Further, we demonstrated SNR evaluation over an 80~km SSFM single-span transmission using 589$\times$96 GBaud channels, corresponding to almost 59~THz optical bandwidth, fully populating the O, E, S, C, L and U bands (1260$-$1675 nm). The SNR evaluation is completed in just 3.6 seconds using four Nvidia V100 16GB PCIe GPUs. Finally, we used this model to find the optimum launch power profile for this system achieving 747~Tbps of potential throughput over 80~km fibre and demonstrating its suitability for UWB optimisation routines.
\end{abstract}


\section{Introduction}
\IEEEPARstart{I}{n} our increasingly interconnected world, the demand for faster and more reliable data transmission has driven the rapid evolution of optical communication systems. In the past decade, there has been research interest towards increasing optical link capacity by expanding the optical bandwidth beyond the conventional C-band. This has been accelerated by emerging research in optical amplification, such as Bismuth-doped fibre amplifiers (BDFA) capable of amplifying the original~\cite{10117285,elson_96-thz_2024} and extended wavelength (O- and E-) bands, incuding distributed Raman-amplified links~\cite{10343123,10302295}. Recent experimental ultrawideband (UWB) system demonstrations have convincingly shown the practical feasibility of transmitting multiple bands including the O-band, using 25~THz of bandwidth spanning original- to ultralong- (O- to U-) bands~\cite{ECOC23_Soma}. The advances in the expansion of the practically feasible transmission bandwidth have, in turn, highlighted the urgent need for rapid modelling and optimisation tools.

\begin{figure}[!t]
\centering
\includegraphics{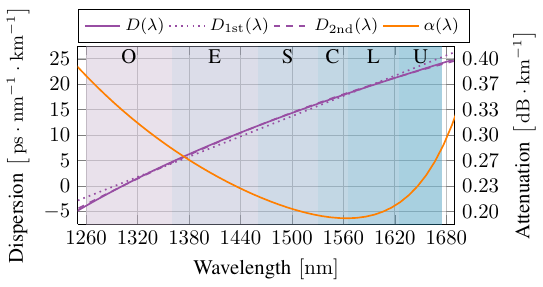}
\vspace{-1.8em}
\caption{Modelled fibre attenuation $\alpha(\lambda)$, dispersion profile $D(\lambda)$ and its fitting curves: $D_\mathrm{1st}(\lambda)$ considering $\beta_2$ and $\beta_3$ terms, and $D_\mathrm{2nd}(\lambda)$ which also includes $\beta_4$ term.}
\label{fig:fibre1}
\vspace{-1em}
\end{figure}

\begin{figure}[!t]
\centering
\includegraphics{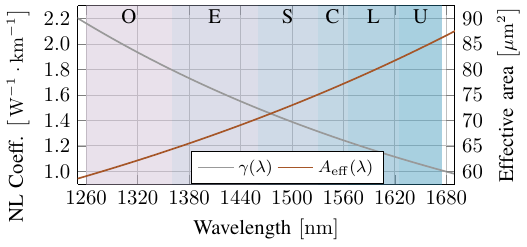}
\vspace{-1.8em}
\caption{Modelled fibre nonlinear (NL) coefficient $\gamma(\lambda)$ and effective area $A_\mathrm{eff}(\lambda)$.}
\label{fig:fibre2}
\vspace{-1em}
\end{figure}

The modelling of UWB systems enables to optimise the design of a given optical communication link to achieve maximum throughput, including finding optimal signal launch power, data rates, modulation formats, number of channels, and, in the case of hybrid-amplified links, also the Raman amplifier's pump powers and frequencies. The Gaussian noise closed-form model (GN-CFM) expressions~\cite{10250945,carena_egn_2014,semrau_closed-form_2019,semrau_modulation_2020,ferrari_gnpy_2020} have increasingly been used for rapid evaluation of nonlinear interference (NLI) noise, which, together with amplified spontaneous emission (ASE) noise and transceiver noise, define the received SNR. GN-CFM, in turn, can be used to optimise optical link parameters. When considering wavelength-dependent fibre parameters, the Gaussian noise (GN) model has been used to evaluate the capacity of UWB systems~\cite{shevchenko_maximizing_2022,poggiolini_closed_2022,damico_scalable_2022}. In most of these studies however, the O-Band is omitted or a guard-band is placed in the zero dispersion region~\cite{ferrari_assessment_2020,hoshida_ultrawideband_2022}, thus, not assessing the potential capacity increase brought by the channels in this band. The guard-band is placed as GN-CFM expressions lack multichannel interference (MCI) expressions arising from four-wave mixing, causing them to underestimate NLI in the case of low chromatic dispersion. This is especially critical for channels around the zero dispersion wavelength of the fibre.
The GN-CFM in \cite{zefreh_accurate_2021} includes MCI expressions, making it valid for multi-channel simulations at and close to zero-dispersion wavelength; however, it does not include inter-channel stimulated Raman scattering (ISRS) terms, making it unsuitable for UWB and scenarios with distributed and hybrid amplification technologies.

In this work, we present, for the first time, an ISRS Integral GN model for UWB transmission including the O-band and channels covering the zero-dispersion wavelength. We have validated the accuracy of this model at zero-dispersion wavelength with a split-step Fourier Method (SSFM) from 1 to 101 channel transmission. Furthermore, we extended the GN model to include the higher-order dispersion terms of the wavelength-dependent parameters to increase accuracy covering over 415~nm modulated bandwidth. Finally, we have implemented a fast numerical integral GN model on multiple GPUs to demonstrate a 589$\times$96~GBaud channels transmission launch power optimisation. This is the first model predicting the performance of a fully loaded O-to-U band transmission without guard bands at zero-dispersion wavelength, which is also capable of estimating the quality of transmission for distributed Raman and hybrid amplification.

The rest of the paper is as follows. In Section~\ref{sec:fibre}, we estimate wavelength-dependent parameters suitable for UWB simulation by modelling standard single-mode fibre (SSMF). In Section~\ref{sec:implementation}, we present a parallelised and GPU-accelerated implementation of the numerical GN integral model
GN model NLI results are compared with the SSFM simulation, considering transmission in the O-band around the zero-dispersion wavelength. In Section \ref{sec:optimisation}, we used the implementation of the GN integral model to optimise the launch power in the UWB transmission system. Appendix~\ref{sec:gn_model} describes the mathematical framework for the integral GN model. Appendix~\ref{sec:closed_form} presents the higher dispersion order expressions for the GN-CFM.

\section{Ultrawideband Fibre Model}\label{sec:fibre}

This section describes the fibre model, which was used to obtain wavelength-dependent parameters suitable for UWB transmission simulations. The contribution of the higher-order dispersion, nonlinear coefficient ($\gamma$) and effective core area (\Aeff) terms becomes non-negligible when increasing total signal optical bandwidth. As $\gamma(f)$ and $A_\mathrm{eff}(f)$ are challenging to measure experimentally over a very large bandwidth, we estimate them by employing a mode solver exploring the azimuthal symmetry of optical fibres~\cite{ferreira_mode_2022}. The mode solver models electromagnetic surface waves along a dielectric waveguide using Maxwell's equations. Unknown propagation coefficients are obtained by numerically solving differential equations with imposed boundary conditions \cite{dil_propagation_nodate}.

In our study, we consider a single-step index fibre with 8.2~$\mathrm{\mu m}$ core diameter and relative core-cladding index difference $\Delta = 0.36\%$. We assume no cladding trench or refractive index gradient. The fibre is modelled to be ITU-T G.652.D standard compliant. The results of the mode solver are shown in Fig.~\ref{fig:fibre1} and \ref{fig:fibre2}.

The fibre attenuation profile shown in Fig.~\ref{fig:fibre1} is modelled considering the Rayleigh scattering and the infrared absorption \cite[Eq.~(1)]{shevchenko_maximizing_2022} to match commercially available SSMFs. The fibre is assumed to not have $\mathrm{OH}^{-1}$ ions absorption or other peaks. Bending loss is assumed to be negligible.

The total dispersion $D$ obtained by the mode solver is shown in Fig.~\ref{fig:fibre1}. $D$ is fit using first ($D_\mathrm{1st}$) and second ($D_\mathrm{2nd}$) order polynomials. Polynomial coefficients directly correspond to the chromatic dispersion parameter $D$, dispersion slope $S$ and dispersion curvature $\dot{S}$. We can find second ($\beta_2$), third ($\beta_3$) and fourth ($\beta_4$) order group velocity dispersion parameters by using \eqref{eq:d}, \eqref{eq:d_slope} and \eqref{eq:d_curve}, respectively:
\begin{align}
\beta_2(\lambda) &= - \frac{D \lambda^2}{2 \pi c}\label{eq:d},\\
S \triangleq \left(\frac{\partial D}{\partial\lambda}\right)_{\lambda=\lambda_{c}}, ~\beta_3(\lambda) &=  \frac{\lambda^3}{(2 \pi c)^2} \left(2 D + S \lambda \right)\label{eq:d_slope},\\
\dot{S} \triangleq \left(\frac{\partial S}{\partial\lambda}\right)_{\lambda=\lambda_{c}}, ~\beta_4(\lambda) &= - \frac{\lambda^4}{(2 \pi c)^3} \left(6 D + 6 S \lambda + \dot S \lambda^2 \right),
    \label{eq:d_curve}
\end{align}

where $c$ is the speed of light in vacuum, $\lambda$ is wavelength and $\lambda_c$ is reference wavelength (centre of modulated signal bandwidth). The average error over the wavelength range of 1260$-$1675~${\rm nm}$ between the $D$ and $D_\mathrm{1st}$ is 0.58~${\rm ps}\cdot{\rm nm}^{-1}\cdot{\rm km}^{-1}$ with a maximum error of 1.7~${\rm ps}\cdot{\rm nm}^{-1}\cdot{\rm km}^{-1}$ at 1260~${\rm nm}$. Modelled $D$ zero-dispersion wavelength $\lambda_{\rm zd}=1302.3~{\rm nm}$ which is within ITU-T G.652.D specification. The $D_\mathrm{1st}$ approximation has zero-dispersion at 1293~${\rm nm}$, which is a large 9.3~${\rm nm}$ error. With a higher-order $\beta_4$ term, the average error between the $D$ and $D_\mathrm{2nd}$ is 0.089~${\rm ps}\cdot{\rm nm}^{-1}\cdot{\rm km}^{-1}$ with a maximum of 0.28~${\rm ps}\cdot{\rm nm}^{-1}\cdot{\rm km}^{-1}$ at 1260~nm. Additionally, including $\beta_4$ term results in only 0.05~${\rm nm}$ error between $\lambda_\mathrm{zd}$ given by the fitted polynomial and the actual $D$ profile.

The wavelength-dependent effective core area $\Aeff$ and the nonlinear coefficient $\gamma$ are shown in Fig.~\ref{fig:fibre2}. $\gamma$ depends on two wavelength-dependent components -- the $\Aeff$ and the nonlinear (Kerr) refractive index $n_2$, shown in \eqref{eq:gamma}:
\begin{equation}
    \gamma(\lambda) = \frac{2\pi}{\lambda} \cdot \frac{n_2(\lambda)}{\Aeff(\lambda)}.
    \label{eq:gamma}
\end{equation}
The nonlinear refractive index depends on two other wavelength-dependent components as shown in \eqref{eq:n2}. The first one is the linear refractive index $n_0$, which we acquire from the mode solver.
The second one is the real component of the third-order optical susceptibility, which is a fourth-rank tensor denoted as $\chi_{xxxx}^{(3)}$. Its value is also wavelength dependent, we consider a linear fit based on two points reported in \cite{PhysRevB.61.10702}.
\begin{equation}
    n_2(\lambda) = \frac{3}{4 \varepsilon_0 c n_0^2 (\lambda)} \, \Re \left\{ \chi_{xxxx}^{(3)} \right\}\,,
    \label{eq:n2}
\end{equation}
where $\varepsilon_0$ is the vacuum permittivity.
Intermodal crosstalk due to coupling between the fundamental and higher-order fibre modes is assumed to be negligible.

For the following simulations, we use the measured ITU-T G652.D fibre Raman gain profile~\cite[Fig. 1b]{10250945} and scaled at different wavelengths by the value~$\Aeff(\lambda)$.

\section{Integral GN Model Implementation}\label{sec:implementation}

In this section, we describe the numerical GN model implementation, optimal computation parameters, and its accuracy at zero-dispersion wavelength. We based it on the ISRS GN integral model introduced in~\cite{semrau_gaussian_2018}. The model is used for the per-channel NLI estimation and is valid for signals with Gaussian-distributed constellations.
We extend the ISRS GN integral model to include frequency-dependant nonlinear coefficient ($\gamma$), and effective core area.

It is further expanded by assuming that the optical power along fibre distance is slowly varying and can be approximated by dividing it into discrete distance steps $N_M$ as shown in \eqref{eq:slow_vary}. The new model is shown in \eqref{eq:new_gn_model}, where the power spectral density (PSD) of NLI noise $G(z,f)$, $L_k$ is the length of fibre span $k$, $G_\mathrm{Tx}(f)$ is signal PSD which we apprimate as rectangular function for the channel bandwidth, $p_k$ is the power profile term shown in \eqref{eq:pk}, which is derived by solving Raman power evolution ordinary differential equations (ODEs), including $\Aeff(f)$. The phase mismatch factor $\phi$ is extended to include a higher-order $\beta_4$ dispersion in \eqref{eq:phi}. We define the Fourier transform as $\mathfrak{F}\{f(x)\}(t) \triangleq \int_{-\infty}^{+\infty} f(x) e^{-j2\pi t x} dx$ and $\sinc\left(x\right) \triangleq \frac{\sin\left(x\right)}{x}$. The mathematical derivations can be found in Appendix~\ref{sec:gn_model}.
\begin{align} \nonumber
    G(z,f) &\approx \frac{16}{27} \gamma^2(f) \int df_1 \int df_2 ~ G_\mathrm{Tx}(f_1) \, G_\mathrm{Tx}(f_2) \\
    &\cdot G_\mathrm{Tx}(f_1 + f_2 - f) \cdot \biggl | \,\sum_{k=1}^{n} \sum_{\forall m} \, p_k(f_1,f_2,\tilde{z}_m) \label{eq:new_gn_model}\\
    &\cdot \Delta z_m \, e^{j\phi(f_1,f_2)(\tilde{L}_k + z_m)} \,  \sinc \! \left(\phi(f_1,f_2) \frac{\Delta z_m}{2}\right) \biggr |^2,\nonumber
\end{align}
\vspace{-0.8em}

\begin{figure}[!t]
\centering
\includegraphics[width=\linewidth]{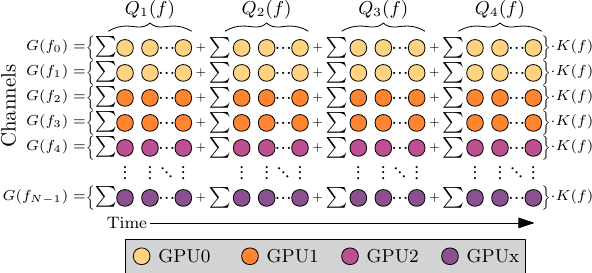}
\vspace{-1em}
\caption{A diagram for the parallel computation progress. Rows represent each channel GN integral NLI solution $G(f)$ using Riemann sum. Each coloured circle represents the computation of a single Riemann sum step for a single quadrant $Q_\kappa$ with $N_R$ samples per quadrant. $K$ represents channel constants outside the integral.}
\label{fig:gpu_nodes}
\vspace{-1em}
\end{figure}

The frequency integrals are further transformed to four quadrants $Q_\kappa$ in the hyperbolic formulation in \eqref{eq:q1} to \eqref{eq:q4}, which increases the rate of convergence, thus speeding up the numerical integration. For numerical implementation, we use Riemann sum to solve $Q_\kappa$. NLI evaluation for every channel is highly parallelisable as NLI computation is independent for each channel. Channels are organised into groups equally distributed to the available GPUs for computation of their respective NLI (each group represented in different colours in Fig.~\ref{fig:gpu_nodes}). Each GPU concurrently computes multiple channels with a batch, one Riemann sum term at a time (each circle in Fig.~\ref{fig:gpu_nodes}), while at the same time, all GPUs compute their channel batches in parallel. Each integral quadrant ($Q_\kappa$) is computed one after another on the same GPU. This approach maximises each GPU utilisation, significantly reducing computation time. The implementation allows the utilisation of an arbitrary number of GPUs in parallel from 1 up to the number of channels ($N_\mathrm{ch}$), with a higher number of GPUs giving diminishing returns as each GPU becomes underutilised.

The code was implemented in Python with JAX
framework~\cite{schoenholz_jax_2021}, and all operations were performed using double-precision (64~bit) floating-point numbers.

\subsection{Numerical model parameters}\label{subsec:hyperp}

\begin{figure}[!b]
\vspace{-1.8em}
\centering
\includegraphics{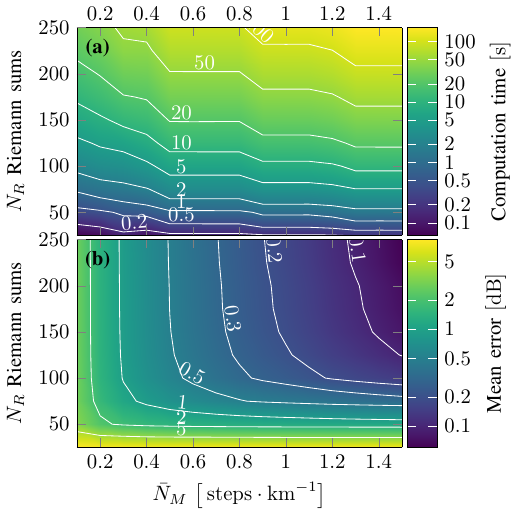}
\vspace{-2.5em}
\caption{(a) Computation time and (b) accuracy versus numerical GN integral model parameters.}
\label{fig:hyperp1}
\end{figure}

In this subsection, our objective is to investigate the accuracy of the NLI of the GN integral model and the computation time trade-offs with different values of the number of Riemann samples ($N_R$); and distance steps $N_M$ along the fibre length. Steps $N_M$ are sampled in log-scale distance intervals, with more steps at the start of the fibre where the highest optical power is. The mean number of steps per km is defined as $\bar{N}_M=(N_M-1)/L\cdot10^3$ where $L$ is fibre length.

We consider the UWB transmission scenario described in Table~\ref{tab:parameters} with the fibre parameters from Fig.~\ref{fig:fibre1} and Fig.~\ref{fig:fibre2}. We use spectrally uniform launch power of 2~dBm per channel. We use nonlinear interference in decibels $\eta_\mathrm{NLI}^\mathrm{dB} \triangleq 10\log_{10}(\eta_\mathrm{NLI})$ as model accuracy metric. We start NLI evaluation $\eta_\mathrm{NLI,0}^\mathrm{dB}$ with large values of $N_R=500$  and $\bar{N}_M=2$ to ensure high accuracy results. Next, we evaluate $\eta_\mathrm{NLI,x}^\mathrm{dB}$ for a number of $\mathrm{x}$ scenarios sweeping through $N_R$ ranging from $25$ to $250$ and $\bar{N}_M$ ranging from $0.1$ to $1.5$ steps/km. Fig.~\ref{fig:hyperp1}a shows the computation time in seconds for NLI estimation obtained with the different values of $N_R$ and $N_M$ using four Nvidia V100 16GB PCIe GPUs. The difference between high accuracy NLI results and swept results $\left| \eta_\mathrm{NLI,0}^\mathrm{dB} - \eta_\mathrm{NLI,x}^\mathrm{dB} \right|$ are shown in Fig.~\ref{fig:hyperp1}b. The results show a trade-off between speed and accuracy is in favour of larger $\bar{N}_M$ rather than large $N_R$ values. For cases where $N_R=75$, $\bar{N}_M=0.95$, taking 3.6~seconds to compute resulting in 0.46~dB NLI error; $N_R=150$, $\bar{N}_M=1.4$ taking 37~seconds and resulting in $<0.1$~dB NLI error. These parameter values can be selected depending on the desired accuracy and execution time.

\subsection{Model accuracy at zero dispersion}\label{subsec:zero_disp}

\begin{figure*}[!t]
\centering
\includegraphics{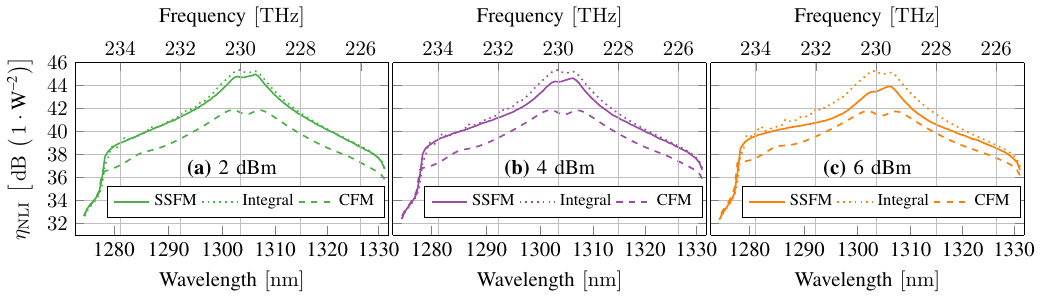}
\vspace{-1.2em}
\caption{Nonlinear interference ($\eta_\mathrm{NLI}$) solutions for 101 channel transmission centered at $\lambda_{\rm zd}$. The figure compares SSFM, Integral, and GN CFM models. Only three cases are shown for 2~dBm, 4~dBm and 6~dBm launch power per channel. The optimum power in this scenario is 0.25~dBm per channel.}
\label{fig:ssfm}
\vspace{-1em}
\end{figure*}

The accuracy of the proposed integral GN model is validated around the zero dispersion wavelength of the fibre through comparisons with SSFM simulations. It is infeasible to simulate a transmission occupying O-to-U bands as it requires a large number of channels which makes SSFM computation expensive. Therefore, here, we limit our accuracy comparison to up to 101 channels. GN integral model has previously been validated for S, C and L bands~\cite{semrau_isrs_2018,10484692,10250945}, therefore we simulate in the O-band, at the zero-dispersion wavelength i.e. at $\lambda_{\rm zd}=1302.3~$nm, where the highest error is expected due to dispersion being a dominant term in the interplay between phase mismatch factor ($\phi$), $\gamma$ and ISRS effects. O-band validation simulation includes the wavelength-dependent fibre parameters as described in section~\ref{sec:fibre}, except that we considered a constant nonlinear coefficient of $\gamma(\lambda_{\rm zd}) \simeq 2$~${\rm W}^{-1} \cdot {\rm km}^{-1}$~-- this is because we do not include frequency-dependent $\gamma$ in the SSFM algorithm. We consider the channel spacing and symbol rate shown in Table~\ref{tab:parameters}, transmitted over 80~km single fibre span. We simulated ten scenarios with the number of channels ranging from 1 to 101, centred at the $\lambda_{\rm zd}$. Each of these scenarios is simulated with five different launch powers from \mbox{--2~dBm} to 6~dBm per channel. The SSFM simulation considers $2^{16}$ random Gaussian symbols per polarisation per channel with two samples per symbol and a root-raised-cosine filter with a roll-off factor of $1\%$. Step sizes for the SSFM simulation were optimised using the local-error method~\cite{sinkin_optimization_2003} using a small goal error value of $\delta_G=10^{-10}$ to ensure accurate results. The integral model was solved using $\bar{N}_M=1.4$ steps and $N_R = 150$, based on results found in Subsection~\ref{subsec:hyperp}.

\begin{figure}[!t]
\centering
\includegraphics{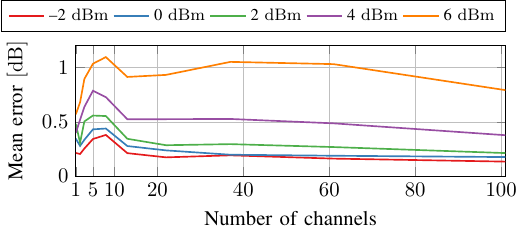}
\vspace{-1.8em}
\caption{Comparing difference between SSFM and GN model nonlinear interference ($\eta_\mathrm{NLI}$) in O-Band.}
\label{fig:errors}
\vspace{-2em}
\end{figure}

For the accuracy metric, we define the channel-wise error between the integral GN model and the SSMF as $\left| \eta_\mathrm{NLI,GN}^\mathrm{dB} - \eta_\mathrm{NLI,SSFM}^\mathrm{dB} \right|$. We evaluate this error for ten different channels and five different launch powers, with the mean channel error shown in Fig.~\ref{fig:errors}. The mean error decreases when increasing the number of channels, the worst scenario being at 6~dBm launch power with a maximum error of 1.1~dB with 8 channels, decreasing to 0.79~dB error at 101 channels. The same trend of mean error peaking at 5 to 8 channels is observed in other scenarios as the maximum error close to zero dispersion channel increases up to a few channels and then remains relatively constant. Channels further from zero dispersion have a better agreement - therefore decreasing mean error with more added channels.

We plot NLI per channel for the 101~channel scenario case in Fig.~\ref{fig:ssfm}. The mean and maximum channel error, together with the total optical power are shown in Table~\ref{tab:errors}. The maximum NLI error is 1.8~dB in the channel with the highest NLI at 1303.4~nm being for the case with 6~dBm launch power per channel shown in Fig.~\ref{fig:ssfm}c. This demonstrates an extreme case where lauch power is much grather than optimum as discussed later in this section. The maximum NLI error is followed by 1~dB at the same wavelength for the scenario with 4~dBm per channel in Fig.~\ref{fig:ssfm}b and 0.7~dB error at 1277.9~nm with 2~dBm per channel in Fig.~\ref{fig:ssfm}a. For lower launch powers, the maximum NLI error per channel is below 0.6~dB. For all the cases good agreement is found between integral and SSFM models in channels that are located further away from the highest NLI point as can be seen in channels below 1292~nm and above 1315~nm in Fig.~\ref{fig:ssfm}b and channels below 1287~nm and above 1321~nm in Fig.~\ref{fig:ssfm}c. Results for optical powers below 2~dBm are very similar to 2~dBm case and, therefore, not included. The error between the GN model and the SSFM is expected to be larger for high launch powers because of the first-order regular perturbation assumption underlying the derivation of the GN model, which results in a NLI overestimation.

\begin{table}[!t]
\begin{center}
\caption{$\eta_\mathrm{NLI}$ error between SSFM, Numerical GN integral model and GN Closed-Form model in 101~channel scenario}
\label{tab:errors}

\bgroup
\setlength{\tabcolsep}{0.4em}
\begin{tabular}{|r|c|c|c|c|c|c|}\hline
Ch. Launch power  $\left[{\rm dBm}\right]$ &
\cellcolor{Set1-A!30}--2 &
\cellcolor{Set1-B!30}0 &
\cellcolor{gray!30}0.25 &
\cellcolor{Set1-C!30}2 &
\cellcolor{Set1-D!30}4 &
\cellcolor{Set1-E!30}6
\\\hline
Tot. Launch power  $\left[{\rm dBm}\right]$ & 18 & 20 & 20.3 & 22 & 24 & 26 \\\hline
SSFM vs. GN-Int. (Mean) $\left[{\rm dB}\right]$ & 0.13 & 0.18 & 0.18 & 0.21 & 0.38 & 0.79 \\\hline
SSFM vs. GN-Int. (Max.) $\left[{\rm dB}\right]$ & 0.65 & 0.59 & 0.60 & 0.72 & 1.02 & 1.81\\\hline
SSFM vs. GN-CFM (Mean) $\left[{\rm dB}\right]$ & 1.84 & 1.78 & 1.78 & 1.75 & 1.58 & 1.14 \\\hline
SSFM vs. GN-CFM (Max.) $\left[{\rm dB}\right]$ & 3.47 & 3.39 & 3.38 & 3.28 & 2.99 & 2.37 \\\hline
\end{tabular}
\egroup
\vspace*{-1.8em}
\end{center}
\end{table}

\begin{figure*}[!b]
\vspace{-1em}
\centering
\includegraphics{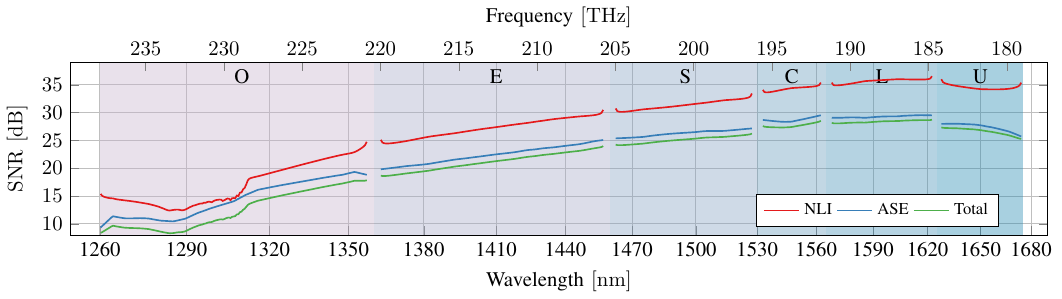}
\vspace{-1.2em}
\caption{SNR result after launch power optimisation for single 80~km fibre span, including nonlinear interference (NLI) and ASE noise. NLI is evaluated with a numerical GN integral model, including $\Aeff(f)$, $\gamma(f)$, $\beta_2(f)$, $\beta_3(f)$ and $\beta_4(f)$ terms.}
\label{fig:SNR}
\end{figure*}

For the same 101~channel scenario, we also include, for comparison, the results obtained using the GN-CFM~\cite{semrau_closed-form_2019}, which were expanded to support the higher order $\beta_4$ dispersion term, as described in the Appendix \ref{sec:closed_form}.
Estimation of NLI using GN-CFM is shown in Fig.~\ref{fig:ssfm}. The average and maximum channel error of $\eta_\mathrm{NLI}^\mathrm{dB}$ is shown in Table~\ref{tab:errors}. The large error is expected because GN-CFM lacks MCI expressions, making it underestimate NLI. The results show that the NLI error decreases with higher launch power, but this is simply because the NLI is overestimated due to the first-order regular perturbation assumption, the same as seen with GN integral model, combined with the NLI underestimation due to lack of MCI expressions.

Additionally, the optimisation algorithm of Section~\ref{sec:optimisation} is used to find the spectrally uniform optimal launch power for the 101~channel scenario described in this section, considering a uniform 7~dB amplifier noise figure. This resulted in 0.254~dBm per channel which is included in Table~\ref{tab:errors}. The integral GN model shows good agreement with SSFM NLI results at optimal launch power with 0.18~dB mean error as well as in lower and higher launch powers up to 4~dBm.

\section{Launch Power Optimisation}\label{sec:optimisation}

In this section, we demonstrate the practical application of the fast numerical ISRS GN model evaluation on a GPU by optimising channel launch powers to maximise total data throughput. The transmission scenario is outlined in Table~\ref{tab:parameters} and optimisation is performed by using the L-BFGS-B algorithm \cite{2020SciPy-NMeth}. This algorithm was chosen because it efficiently finds a local minimum with a large problem dimensionality, with finite-differences gradient estimation. To simplify the problem, we reduced the number of optimisation dimensions by dividing the spectral window into segments for power optimisation. The number of segments per band is defined as $N_B = \mathrm{round} ( B_\mathrm{band} / B_p )$ , where $B_\mathrm{band}$ is the total bandwidth of each band and $B_p$ approximate segment bandwidth, which value we chose as 750~GHz in the case of O-Band and 1.5~THz for all other bands. These segments are equally spaced within the band with a minimum of two edges per band. We optimise segment edges, whose values are represented in the dBm scale, as this allows the optimiser to effectively search for positive and negative values in the logarithmic domain. The optimisation was bounded from --5~dBm to 5~dBm. Channel powers were linearly interpolated between the segments' edges, enabling the optimiser to evaluate more complex power profiles for each band. Despite that the L-BFGS-B algorithm does not guarantee to find the global minimum solution, we consider the solution to be approaching towards the optimum, considering the practical application~\cite{ECOC23_Vasylchenkova}.

\begin{table}[!t]
\begin{center}
\caption{Simulation parameters}
\vspace*{-1em}
\label{tab:parameters}
\begin{tabular}{| c | cl |}\hline
\# of channels ($N_\mathrm{ch}$) & 589 & \\\hline
\# of guard channels ($N_\mathrm{A}$) & 32 & \\\hline
Modulation format & Gaussian & \\\hline
Symbol rate & 96 & ${\rm GBaud}$ \\\hline
Channel spacing & 100 & ${\rm GHz}$ \\\hline
Reference wavelength ($\lambda_c$) & 1438 & ${\rm nm}$ \\\hline
Fibre length ($L$) & 80 & ${\rm km}$ \\\hline
Dispersion ($D$) & 17.74 & ${\rm ps}\cdot {\rm nm}^{-1} \cdot {\rm km}^{-1}$ \\\hline
Dispersion slope ($S$) & 0.057 & ${\rm ps}\cdot {\rm nm}^{-2} \cdot {\rm km}^{-1}$ \\\hline
Dispersion curvature ($\dot S$) & --5.975$\cdot \text{10}^\text{-5}$ & ${\rm ps}\cdot {\rm nm}^{-3} \cdot {\rm km}^{-1}$ \\\hline
Guard band gap & 5 & ${\rm nm}$ \\\hline
\end{tabular}
\vspace*{-2.5em}
\end{center}
\end{table}

\begin{table}[!t]
\begin{center}
\caption{Per band characteristics}
\vspace*{-1em}
\label{tab:nf}
\begin{tabular}{|c|c|c|c|c|c|c|}\hline
Band &
\cellcolor{grad-1!50}O & \cellcolor{grad-2!50}E & \cellcolor{grad-3!50}S & \cellcolor{grad-4!50}C & \cellcolor{grad-5!50}L & \cellcolor{grad-6!50}U \\\hline
Amp. noise figure $\left[{\rm dB}\right]$ & 7 & 7 & 7 & 5 & 6 & 8 \\\hline
Num. of channels & 171 & 143 & 88 & 38 & 65 & 52 \\\hline
Num. of opt. segments & 23 & 10 & 6 & 3 & 4 & 3 \\\hline
\end{tabular}
\vspace*{-2.5em}
\end{center}
\end{table}

The cost function is defined as the link throughput shown in \eqref{eq:loss}, where $N_\mathrm{ch}$ is the number of channels and $A$ denotes a subset of all channels that we leave empty, i.e., the set of guard bands. $P_i$~is the channel power and $\eta_{\mathrm{NLI},i}$ is NLI noise factor per channel. $P_{\mathrm{ASE},i}$ is ASE noise variance per channel over both polarisations and we assume all band amplifiers ideally compensate for lost power back to optimised launch power profile.
\vspace{-1em}
\begin{equation}
    \mathcal{L} = -\sum_{i=0,\,i\notin A}^{N_\mathrm{ch}} \log_2 \left ( 1 + \frac{P_i}{\eta_{\mathrm{NLI},i} \cdot P_i^3 + P_{\mathrm{ASE},i}} \right ).
    \label{eq:loss}
\end{equation}
A uniform 0~dBm per band launch power was used as the initial condition. The optimiser is run with stop condition of $\max(\left | \nabla \mathcal{L} \right |) \leqslant 0.01$. The four Nvidia V100 16GB PCIe GPUs completed the optimisation task in 641 seconds using $N_R=75$ and $\bar{N}_M=1$. We then re-ran the optimisation with these results as the initial condition and with higher accuracy parameters of $N_R=150$ and $\bar{N}_M=1.4$ to fine-tune the solution. This took another 72 minutes.

\begin{figure}[!t]
\centering
\includegraphics{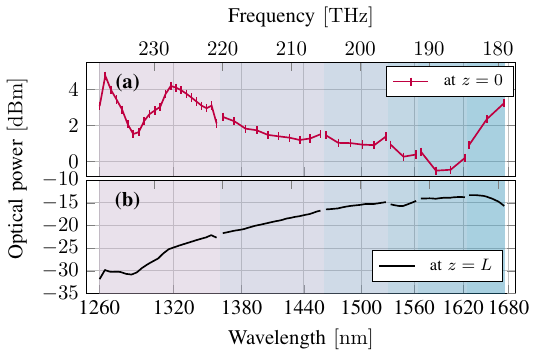}
\vspace{-2.5em}
\caption{Transmitted optimised launch power (a), and optical power at the end of span (b), $L=80~\text{km}$. Each vertical line represents a segment edge which is controlled by the optimiser and channel launch power is linearly interpolated between these edges.}
\label{fig:Power}
\vspace{-1em}
\end{figure}

\begin{table}[!b]
\vspace{-1.8em}
\begin{center}
\caption{Per band characteristics in optimised power segment scenario (Case 1) and spectrally uniform optical launch power (Case 2). "Pow." represents total optical power per band, and "cap." represents potential throughput, * denotes case with non-ideal transceiver noise. }
\vspace*{-1em}
\label{tab:opt}
\bgroup
\setlength{\tabcolsep}{0.65em}
\begin{tabular}{|c|c|c|c|c|c|c|}\hline
Band &
\cellcolor{grad-1!50}O & \cellcolor{grad-2!50}E & \cellcolor{grad-3!50}S & \cellcolor{grad-4!50}C & \cellcolor{grad-5!50}L & \cellcolor{grad-6!50}U \\\hline
Case 1 pow. $\left[{\rm dBm}\right]$ & 25.6 & 23.2 & 20.5 & 16.3 & 18.0 & 19.4 \\\hline
Case 2 pow. $\left[{\rm dBm}\right]$ & 24.2 & 23.4 & 21.3 & 17.7 & 20.0 & 19.0 \\\hline
Case 1 cap. $\left[{\rm Tbps}\right]$ & 139.6 & 193.2 & 141.0 & 67.1 & 117.7 & 88.4 \\\hline
Case 2 cap. $\left[{\rm Tbps}\right]$ & 139.8 & 189.0 & 134.6 & 66.4 & 119.3 & 89.8 \\\hline
Case 1$^{*}$ cap. $\left[{\rm Tbps}\right]$ & 130.4 & 159.9 & 106.0 & 46.9 & 80.7 & 63.7 \\\hline
Case 2$^{*}$ cap. $\left[{\rm Tbps}\right]$ & 130.9 & 158.2 & 104.3 & 46.8 & 80.9 & 63.9 \\\hline
\end{tabular}
\egroup
\vspace{-.8em}
\end{center}
\end{table}

The results of the launch power optimisation are shown in Fig.~\ref{fig:Power}a, together with received power after ISRS effects in Fig.~\ref{fig:Power}b. The total optical power launched into the fibre using optimised parameters is 29.4~dBm, and per-band powers are shown in Table~\ref{tab:opt}. All bands show a simple launch power profile, indicating that the optimisation using the segment scheme is a good approximation, compared to a more expensive per-channel optimisation scheme. E- to L-Bands show an increase in launch power at the edges of each band as these regions have lower NLI due to band guards. O-Band has a more complex profile and this is the reason for using a smaller $B_p$ segment bandwidth. The peak launch power of 4.7~dBm was observed at 1264.5~nm, where fibre loss is highest. The lowest launch power optimisation segment edge of 1.54~dBm is at 1285.7~nm, which indicates the highest NLI region. Comparing this highest NLI region to the model accuracy at zero dispersion results in Subsection~\ref{subsec:zero_disp} we can conclude that the model should be accurate for this power optimisation scenario for the given low launch power in this region.

The SNR results together with NLI and ASE noise are shown in Fig.~\ref{fig:SNR}. The O-Band exhibits the worst performance and noise variation, with the lowest and highest SNR of 8.3~dB and 17.8~dB respectively. C, L and U bands show the lowest channel noise with a mean SNR of 27.7~dB, 28.4~dB and 26.6~dB respectively.

We estimated the potential link throughput at the Shannon rate $C = -2 B_\mathrm{ch} \cdot \mathcal{L}$ with $B_\mathrm{ch}$ being the channel bandwidth; resulting in 747~Tbps when optimised using the power segment method. The capacity per band is shown in Table~\ref{tab:opt} Case~1. We use the same optimisation algorithm to find optimal launch power with all channels having uniform power across all bands shown in Table~\ref{tab:opt} Case~2. This results in 1.87~dBm per channel (29.33~dBm total power) and 739~Tbps of potential throughput, which is only 1.07\% less than Case~1. High SNR values may be currently unachievable due to high baud rate transceivers. Assuming a non-ideal transceiver 20~dB back-to-back SNR noise for each transmission channel, the potential throughput is reduced to 587.5~Tbps for Case~1 and 585.0~Tbps for Case~2. This indicates that having a complex launch power profile (and a large number of power segments) has only marginal potential throughput improvements.

\section{Conclusion}
In this work, we model the ultrawideband optical transmission, covering 1260~nm to 1675~nm, taking into account wavelength-dependent fibre parameters, including dispersion curvature, nonlinear coefficient and effective fibre core area. We have extended the GN integral model to include these parameters and described its solution using the slowly varying power approximation along the fibre length and frequency integral in hyperbolic coordinates. The GN integral was numerically solved using a Riemann sum, and implementation was optimised for fast and efficient parallel GPU computing. Algorithm parameters were investigated, to achieve a good trade-off between accuracy and computation time. This model includes multi-channel interference (four-wave mixing), enabling accurate simulations for O-band transmission around the $\lambda_{\rm zd}$ (zero-dispersion wavelength). This was confirmed through comparisons with SSFM-based modelling, showing good agreement across the O-band centred at $\lambda_{\rm zd}$, with an average error in the NLI of only 0.21~dB with 101~channels at 2~dBm launch power per channel, while a solution obtained using the GN closed-form expressions resulted in mean channel error of 1.75~dB NLI. The average error of the numerical integral NLI solution decreases with a larger number of channels and lower launch powers.

The SNR calculation takes just 3.6 seconds for O-to-U band transmission over 80 km using four Nvidia V100 16GB PCIe GPUs. We have optimised launch power in a fully loaded O-to-U band system, using the L-BFGS-B algorithm. For a link length of 80~km, the optimisation task with our implementation was completed under 83~minutes, using the same four GPUs.

This model can be used to investigate several other parameters, such as channel symbol rate and spacing, distributed Raman amplification pump powers and wavelengths. Future work will include the investigation of numerical accuracy with multiple spans and constellation correction terms. In addition, a more advanced lumped amplifier gain model should be considered to achieve a closer representation of realistic system scenarios.

\section*{Acknowledgments}
This work is partly funded by the EPSRC Programme Grant TRANSNET~(EP/R035342/1), EWOC~(EP/W015714/1), and UKRI Future Leaders Fellowship~(MR/T041218/1). M. Jarmolovi\v{c}ius and H. Buglia are funded by the Microsoft 'Optics for the Cloud' Alliance and a UCL Faculty of Engineering Sciences Studentship.

\section*{Data Availability Statement}
The data that support the figures in this paper are available from the UCL Research Data Repository (DOI:~\href{https://doi.org/10.5522/04/24975612}{10.5522/04/24975612}), hosted by FigShare.

\begin{appendices}
\section{Numerical Gaussian Noise Model}\label{sec:gn_model}

In this section, we describe in detail the mathematical derivation of Gaussian Noise Model improvements to support wavelength-dependent parameters, approximating power as slowly varying along fibre length integral, and integral in hyperbolic coordinates.

We start with the integral expression~\cite[Eq.~(4)]{semrau_gaussian_2018} in \eqref{eq:gain}. We include the sum operation for multi-span systems, with $n$ number of spans. In the case of identical spans, this summation is reduced to the phased-array term shown in~\cite[Eq.~(5)]{semrau_gaussian_2018}. $\rho(z,f)$ is the normalised power profile obtained by numerically solving the Raman power evolution profile~\cite[Eq.~(1)]{10250945} using ODEs, where Raman gain spectrum $g_r(\Delta f)$ is scaled by the effective area at the channel frequency $f$. % , $P_\mathrm{tot}$ is total optical power, $P_k$ is power of channel $k$, $L_\mathrm{eff}$ is effective fibre length,
% We considered the nonlinear coefficient $\gamma$ to be constant within the channel bandwidth. This allows us to approximate the and not include $\gamma(f)$.
% \begin{align}
%     &G(z,f) = \frac{16}{27} \gamma^2(f) \int df_1 \int df_2  \label{eq:gain}\\
%     &\cdot  G_\mathrm{Tx}(f_1) \, G_\mathrm{Tx}(f_2) \, G_\mathrm{Tx}(f_1 + f_2 - f) \cdot \nonumber
%     \biggl | \, \sum_{k=1}^{n} \int_0^{L_k} d \zeta \\
%     &\cdot \sqrt{\frac{\rho(\zeta,f_1)\rho(\zeta,f_2)\rho(\zeta,f_1+f_2-f)}{\rho(\zeta,f)}} \, e^{j\phi(f_1, f_2, f, L_k + \zeta)} \, \biggr |^2. \nonumber
% \end{align}

% \begin{equation}
% \begin{splitfit}
%     & \Psi(f_1,f_2) = G_\mathrm{Tx}(f_1) \,G_\mathrm{Tx}(f_2) \, G_\mathrm{Tx}(f_1 + f_2 - f)  \cdot \sum_{k=1}^{n} \sum_{\forall m} \,
%     \\
%    & p_k(f_1,f_2,\tilde{z}_m) \cdot e^{j\phi(f_1,f_2)(\tilde{L}_k + z_m)} \cdot \Delta z_m \, \sinc \! \left(\phi(f_1,f_2) \frac{\Delta z_m}{2}\right). \label{eq:psi_fin}
\vspace*{-1.5em}
\begin{gather}
\small
\begin{aligned}
    G(z,f) =& \frac{16}{27} \gamma^2(f) \int df_1 \int df_2  \cdot  G_\mathrm{Tx}(f_1) \, G_\mathrm{Tx}(f_2) G_\mathrm{Tx}(f_1  \\%
     + f_2 - f) & \cdot%
    \biggl | \, \sum_{k=1}^{n} \int_0^{L_k} d \zeta%
    \cdot p_k(f_1,f_2,\zeta) \, e^{j\phi(f_1, f_2, f, L_k + \zeta)} \, \biggr |^2,\label{eq:gain}%
\end{aligned}
\end{gather} where we simplify normalised power profile terms into $p_k$:
\begin{equation}
     p_k(f_1,f_2,\zeta) = \sqrt{\frac{\rho(\zeta,f_1)\rho(\zeta,f_2)\rho(\zeta,f_1+f_2-f)}{\rho(\zeta,f)}}\label{eq:pk}.
\end{equation}

% G(z,f) = \frac{16}{27} \gamma(f)^2 \rho(z,f) \int df_1 \int df_2 \cdot \biggl | \sum_{k=1}^{n} \int_0^{L_k} d \zeta \\
% S_k(f_1, f_2, f) \cdot \frac{P_{tot} e^{-\alpha \zeta -P_k C_r L_\mathrm{eff}(f_1 + f_2 - f)}} {\int G_k(\upsilon) e^{-P_k C_r L_\mathrm{eff} \upsilon} d \upsilon} e^{j\phi(f_1, f_2, f, L_k + \zeta)} \biggr |^2.

We approximate the non-linear coefficient $\gamma$ to be constant within the channel bandwidth. Furthermore, we assume that all interfering channels have the same $\gamma$ as the channel itself, as the most NLI contribution for the channel is local and further spaced channels with the largest $\gamma$ error will have the lowest contribution. This assumption allows us to simply keep the $\gamma(f)$ term outside the integral as shown in \eqref{eq:gain}.
% \begin{multline}
%     G(z,f) = \frac{16}{27} \gamma(f)^2 \int df_1 \int df_2 \\
%     \cdot G_\mathrm{Tx}(f_1) G_\mathrm{Tx}(f_2) G_\mathrm{Tx}(f_1 + f_2 - f) \\
%     \cdot \left | \int_0^L d \zeta \sqrt{\frac{\rho(\zeta, f_1) \rho(\zeta, f_2) \rho(\zeta, f_1 + f_2 - f)}{\rho(\zeta, f)}} e^{j\phi(f_1, f_2, f, \zeta)} \right |^2
%     \label{eq:gain}
% \end{multline}

% \begin{multline}
%     G(z,f) \approx \frac{16}{27} \gamma(f)^2 G_1(f) \int df_1 \int df_2 \Psi(f_1, f_2) \label{eq:gain_expanded}
% \end{multline}

% = \frac{16}{27} \gamma(f)^2 G_1(f) \int df_1 \int df_2 \Psi(f_1, f_2)

%We solved full power evolution profile $p_k$ along the length of the fibre link $\zeta$ \cite{zirngibl_analytical_1998}.  This is expressed in Eq.~\eqref{eq:gain2}.

% We define term $\Psi(f_1, f_2)$ in Eq.~\eqref{eq:gain2}:
% \begin{multline}
%     \Psi(f_1, f_2) = \biggl | \sum_{k=1}^{n} \int_0^{L_k}
%     p_k(f_1,f_2,\zeta) \cdot e^{j\phi(f_1,f_2,f,\tilde{L}_k + \zeta)} d \zeta \biggr |^2 \label{eq:gain2},
% \end{multline}

% which rearranges Eq.~\eqref{eq:gain}, yielding Eq.~\eqref{eq:gain2}:
% \begin{align}
%     & G(z, f) = \frac{16}{27} \gamma(f)^2 \int df_1 \int df_2 \label{eq:gain3} \\
%     & \qquad G_\mathrm{Tx}(f_1) G_\mathrm{Tx}(f_2) G_\mathrm{Tx}(f_1 + f_2 - f) \Psi(f_1, f_2) \nonumber.
% \end{align}

The phase mismatch factor $\phi$ is derived by extending dispersion Taylor series to include the fourth-order dispersion coefficient $\beta_4$ in \cite[Eq.~(13)]{semrau_gaussian_2018} and factoring $\phi(f_m,f_k,f_0,\zeta) = \left[\Gamma(\zeta,f_m) + \Gamma^*(\zeta,f_m + f_k - f_0) + \Gamma(\zeta,f_k) - \Gamma(\zeta,f_0)\right] / j$. This results in \eqref{eq:phi}, similar to~\cite[Eq.~(3)]{poggiolini_closed_2022}:
% \vspace{-.5em}
% \begin{multline}
%     \phi(f_1+f,f_2+f,f_i,\zeta) = -4\pi^2 f_1 f_2 \biggl[ \beta_2 + \pi \beta_3(f_1+f_2+2f_i) + \frac{2 \pi^2}{3} \\
%     \cdot \beta_4 \bigl[ f_1^2 + \frac{3}{2} f_1 f_2 + 3 f_1 f_i + f_2^2 + 3 f_2 f_i + 3 f_i^2 \bigr] \biggr] \zeta.
%     \label{eq:phi}
% \end{multline}

\vspace*{-1.5em}
\begin{gather}\label{eq:phi}
\small
\begin{aligned}
&\phi(f_1+f,f_2+f,f_i,\zeta) = -4\pi^2 f_1 f_2 \biggl[ \beta_2 + \pi \beta_3(f_1+f_2+2f_i) \\
 & + \frac{2 \pi^2}{3} \beta_4 \bigl[ f_1^2 + \frac{3}{2} f_1 f_2 + 3 f_1 f_i + f_2^2 + 3 f_2 f_i + 3 f_i^2 \bigr] \biggr] \zeta.\raisetag{3ex}
\end{aligned}
\end{gather}

We assume power $p_k(f_1,f_2,\zeta)$ is slowly varying, therefore it can be approximated with a sum of discrete distance steps $m$ along the fibre length. $m$ values are distributed in logarithmically increasing distance steps to have approximately equal power change per step. The step size must be small enough to capture $\phi$ phase variations to give accurate NLI, which can be seen from Fig.~\ref{fig:hyperp1}b. This is shown in \eqref{eq:slow_vary0} to \eqref{eq:slow_vary}, where $\Delta z_m$ is the step size distribution with $z_m$ midpoints. $\tilde{z}_m$ is the distance at which the power profile is evaluated within each step. In Subsection~\ref{subsec:hyperp} we demonstrate that the error is small given the large enough number $N_M$ of $m$ steps.
\begin{align}
     & \sum_{k=1}^n \int_0^{L_k} p_k(f_1,f_2,\zeta) \cdot e^{j\phi (f_1, f_2, \tilde{L}_k + \zeta)} d\zeta \\
    & \approx \sum_{k=1}^n \sum_{\forall m} p_k (f_1,f_2,\tilde{z}_m) \int_{z_1(m)}^{z_2(m)} e^{j\phi(f_1,f_2,\tilde{L}_k + \zeta)} d \zeta \label{eq:slow_vary0} \\
    & = \sum_{k=1}^n \sum_{\forall m} p_k (f_1,f_2,\tilde{z}_m) \nonumber \\
    & \qquad\qquad \cdot \frac{e^{j\phi[f_1,f_2,\tilde{L}_k + z_2(m)]} - e^{j\phi[f_1,f_2,\tilde{L}_k + z_1(m)]}}{j\phi(f_1,f_2)} \\
    & = \sum_{k=1}^n \sum_{\forall m} p_k (f_1,f_2,\tilde{z}_m) \cdot e^{j\phi(f_1,f_2,\tilde{L}_k + z_m)} \nonumber \\
    & \qquad\qquad \cdot \frac{e^{j \phi(f_1,f_2)\,\Delta z_m / 2} - e^{-j \phi(f_1,f_2) \, \Delta z_m /2}}{j\phi(f_1,f_2)}\\
    & = \sum_{k=1}^{n} \sum_{\forall m}
    p_k(f_1,f_2,\tilde{z}_m) \cdot e^{j\phi(f_1,f_2)(\tilde{L}_k + z_m)} \nonumber \\
    & \qquad\qquad \cdot \Delta z_m \, \sinc \! \left(\phi(f_1,f_2) \frac{\Delta z_m}{2}\right).
    \label{eq:slow_vary}
\end{align}
% \begin{multline}
%     G(z,f) \approx \frac{16}{27} \gamma(f)^2 G_1(f) \int df_1 \int df_2 \cdot \biggl | \sum_{k=1}^{n} \sum_{\forall m} \\
%     p_k(f_1,f_2,\tilde{z}_m) \cdot e^{j\phi(f_1,f_2)(\tilde{L}_k + z_m)} \cdot \frac{2\sin(\phi(f_1,f_2) \Delta z_m / 2)}{\phi(f_1,f_2)} \biggr |^2 = \\
%     = \frac{16}{27} \gamma(f)^2 G_1(f) \int df_1 \int df_2 \Psi(f_1, f_2)    \label{eq:gain_expanded}
% \end{multline}
This approximation in \eqref{eq:slow_vary} is applied to \eqref{eq:gain} yielding our integral model in \eqref{eq:new_gn_model}.
% \begin{multline}
%     G(z,f) \approx \frac{16}{27} \gamma(f)^2 G_1(f) \int df_1 \int df_2 \Psi(f_1, f_2) \label{eq:gain_expanded}
% \end{multline}
The frequency integral is solved for the entire signal bandwidth $\int_{-B}^{+B} d f_1 \int_{-B}^{+B} d f_2 \Psi(f_1, f_2)$ where $\Psi(f_1, f_2)$ is a generic function and $B$ is half of the total bandwidth. The function is recentered at the channel frequency $f$, and split into four quadrants $Q_\kappa$, where $\kappa = \left\{1, 2, 3, 4\right\}$. Integration limits are swapped, resulting in quadrants with limits between $B\pm f$ and $0$ as shown in \eqref{eq:q1a} to \eqref{eq:q4d}:
\begin{align}
    Q_{1} &= \int_0^{B-f} d f_1 \quad\int_{0}^{B-f} d f_2 \quad\Psi(+f_1, +f_2)
    \label{eq:q1a},\\
    Q_{2} &= \int_0^{B+f} d f_1 \quad\int_{0}^{B-f} d f_2 \quad\Psi(-f_1, +f_2)
    \label{eq:q2b},\\
    Q_{3} &= \int_0^{B+f} d f_1 \quad\int_{0}^{B+f} d f_2 \quad\Psi(-f_1, -f_2)
    \label{eq:q3c},\\
    Q_{4} &= \int_0^{B-f} d f_1 \quad\int_{0}^{B+f} d f_2 \quad\Psi(+f_1, -f_2)
    \label{eq:q4d}.
\end{align}
Each quadrant is transformed into hyperbolic coordinates similar to \cite[Section VIII]{poggiolini_gn_2012}; by using $\upsilon_1 = f_1 f_2$ and \mbox{$\upsilon_2=\ln{\sqrt{f_1/f_2}}$}, yielding \eqref{eq:q1} to \eqref{eq:q4}:
% \vspace{-.5em}
\begin{equation}
    Q_{1} = \int_0^{(B-f)^2} d \upsilon_1\int_{
    -\ln{\left(\frac{B-f}{\sqrt{\upsilon_1}}\right)}}^{
    +\ln{\left(\frac{B-f}{\sqrt{\upsilon_1}}\right)}}
    d \upsilon_2 \Psi(g_1, g_2)
    \label{eq:q1},
\end{equation}
\begin{equation}
    Q_{2} = \int_0^{(B+f)(B-f)} d \upsilon_1\int_{
    -\ln{\left(\frac{B-f}{\sqrt{\upsilon_1}}\right)}}^{
    +\ln{\left(\frac{B+f}{\sqrt{\upsilon_1}}\right)}}
    d \upsilon_2 \Psi(-g_1, g_2)
    \label{eq:q2},
\end{equation}
\begin{equation}
    Q_{3} = \int_0^{(B+f)^2} d \upsilon_1\int_{
    -\ln{\left(\frac{B+f}{\sqrt{\upsilon_1}}\right)}}^{
    +\ln{\left(\frac{B+f}{\sqrt{\upsilon_1}}\right)}}
    d \upsilon_2 \Psi(-g_1, -g_2)
    \label{eq:q3},
\end{equation}
\begin{equation}
    Q_{4} = \int_0^{(B-f)(B+f)} d \upsilon_1\int_{
    -\ln{\left(\frac{B+f}{\sqrt{\upsilon_1}}\right)}}^{
    +\ln{\left(\frac{B-f}{\sqrt{\upsilon_1}}\right)}}
    d \upsilon_2 \Psi(g_1, -g_2)
    \label{eq:q4},
\end{equation}
where $g_1 = \sqrt{\upsilon_1} e^{\upsilon_2}$ and $g_2 = \sqrt{\upsilon_1} e^{-\upsilon_2}$. Combining quadrants with integration with hyperbolic coordinates and slowing varying power approximation, we obtain the final expression in \eqref{eq:gain_fin} and \eqref{eq:psi_fin}:
\vspace*{-0.5em}
\begin{equation}
    G(z,f) = \frac{16}{27} \gamma^2(f) \sum_{\forall \kappa} Q_\kappa(f)
    \label{eq:gain_fin},
\end{equation}
\vspace*{-3em}

\begin{gather}
\small
\begin{aligned}
    & \Psi(f_1,f_2) = G_\mathrm{Tx}(f_1) \,G_\mathrm{Tx}(f_2) \, G_\mathrm{Tx}(f_1 + f_2 - f)  \cdot \sum_{k=1}^{n} \sum_{\forall m} \,
    \\
   & p_k(f_1,f_2,\tilde{z}_m) \cdot e^{j\phi(f_1,f_2)(\tilde{L}_k + z_m)} \cdot \Delta z_m \, \sinc \! \left(\phi(f_1,f_2) \frac{\Delta z_m}{2}\right).\label{eq:psi_fin}
\end{aligned}
\end{gather}
\vspace*{-2em}

\section{Closed-Form Gaussian Noise Model}\label{sec:closed_form}

To include higher order dispersion terms in closed-form expression, we have to extend cross-phase modulation (XPM) contribution phase mismatch term $ \phi_{i,k}$~\cite[Eq.~(34)]{10250945} and self-phase modulation (SPM) contribution phase mismatch term $\phi_{i}$~\cite[Eq.~(42)]{10250945} in \eqref{eq:phi_xpm} and~\eqref{eq:phi_spm}, respectively. In both cases, we assume that the dispersion slope $\beta_3$ and the dispersion curve $\beta_4$ are constant over the channel bandwidth.

\vspace*{-1.5em}
\begin{equation}
\begin{splitfit}
&\phi(f_1 + f_i, f_2 + f_k, f_i)= -4\pi^2 f_1 \Delta f \biggl[ \beta_2 + \pi \beta_3 (f_1 + f_2 + f_i + f_k) + \frac{\pi^{2} \beta_{4}}{3} \cdot \\
 & ( 2 f_{1}^{2} + 3 f_{1} f_{2} + 3 f_{1} f_{i} + 3 f_{1} f_{k} + 2 f_{2}^{2} +  2 f_{2} f_{i} + 4 f_{2} f_{k} + 2 f_{i}^{2} + 2 f_{i} f_{k} + 2 f_{k}^{2} ) \biggr] \\
 &\qquad \approx -4\pi^2 \left [ \beta_2 + \pi \beta_3 (f_i + f_k) + \frac{2\pi^{2} \beta_{4}}{3} ( f_i^2 + f_i f_k + f_k^2) \right ] \cdot (f_k - f_i) f_1  \\
 &\qquad  =\phi_{i,k} \cdot f_1. \label{eq:phi_xpm}\end{splitfit}\raisetag{3\baselineskip}
\end{equation}
\vspace*{-.3em}
SPM phase mismatch has the same derivation as \eqref{eq:phi}.
\begin{align}
\phi(f_1 + f_i, f_2 + f_i, f_i) &\approx -4\pi^2 f_1 f_2 \left [ \beta_2 + 2 \pi \beta_3 f_i + 2 \pi^2 \beta_4 f_i^2 \right ] \nonumber\\
&= \phi_{i} \cdot f_1 f_2.
\label{eq:phi_spm}
\end{align}
\vspace*{-2em}
\end{appendices}
\bibliographystyle{ieeetr}

\end{document}